\documentclass[%
reprint,
superscriptaddress,
showpacs,preprintnumbers,
nofootinbib,
nobibnotes,
 amsmath,amssymb,
aps,
]{revtex4-1}

\usepackage{graphicx}% Include figure files
\usepackage{dcolumn}% Align table columns on decimal point
\usepackage{bm}% bold math
\usepackage{amsmath}
\usepackage{siunitx}
\begin{document}

\title{Backward Propagating Acoustic Waves in Single Gold Nanobeams}
% Force line breaks 
%with \\
\author{Cyril Jean}
\author{Laurent Belliard}\email{Laurent.Belliard@upmc.fr}
\author{Lo\"ic Becerra}
\author{Bernard Perrin}

\affiliation{Sorbonne Universit\'es, UPMC Univ Paris 06, CNRS-UMR 7588, 
Institut des NanoSciences de Paris, F-75005, Paris, France}%


%\affiliation{
%\affiliation{
%\affiliation{%


\date{\today}% It is always \today, today,

\begin{abstract} Femtosecond pump-probe spectroscopy has been carried 
out on suspended gold nanostructures with a rectangular cross section
lithographed on a silicon substrate. With a thickness fixed to 
$\SI{110}{\nano\metre}$ and a width ranging from $\SI{200}{\nano\metre}$ to 
$\SI{800}{\nano\metre}$, size dependent measurements are used to 
distinguish which confined acoustic modes are detected. 
Furthermore, in order to avoid any ambiguity due to the measurement 
uncertainties on both the frequency and size, pump and probe beams 
are also spatially shifted to detect guided acoustic phonons. 
This leads us to the observation of backward propagating acoustic 
phonons in the gigahertz range ($\SI{\sim 3}{\giga\hertz}$) 
in such nanostructures. While backward wave propagation in elastic 
waveguides has been predicted and already observed at the macroscale, 
very few studies have been done at the nanoscale. 
Here, we show that these backward waves can be used as the unique 
signature of the width dilatational acoustic mode.

%\begin{description}
%use the optional argument of the \verb+\item+ 
%command to give the category of each item. 
%\end{description}
\end{abstract}

%\pacs{43.35.+d,62.65.+k,78.20.hc}% PACS, the Physics and Astronomy
                             % Classification figure.
\maketitle

%\tableofcontents


Probing the elasticity at the nanoscale is a challenge that led 
a wide community to study confined acoustic modes of nano-objects
in the $\SI{10}{\giga\hertz}$-$\SI{1}{\tera\hertz}$ range using 
time-resolved pump-probe experiments\cite{crut2015acoustic}. In such an 
approach, acoustic modes are excited by the absorption of a femtosecond 
laser pulse and detected in transmission\cite{Juve2010,VanDijk2005} or reflectivity 
geometry in near\cite{siry2003picosecond,Vertikov1996} or 
far\cite{guillet2009optoacoustic,Bienville2006} field by the 
induced change in the material optical properties. As ensemble 
studies result in inhomogeneous broadening of the acoustic 
features\cite{Kelf2011,del1999coherent}, single particle spectroscopy 
has been considered to circumvent 
this drawback and to clarify the acoustic 
response\cite{Staleva,Zijlstra2008,VanDijk2005}. However, the strong 
damping at these extremely high frequencies has led several groups to 
isolate the nanoresonators from their substrate to avoid 
energy leaking through the nano-object-substrate 
contact\cite{Major2013,Belliard2013,Ristow2013}. 
These breakthroughs have made possible 
the observation of an other source of acoustic energy leaks. Indeed, 
it has been proven that acoustic phonons are also guided along the 
nano-objects\cite{jean2014direct,Mante2013}. Other major developments 
illustrate the possibility to use nano-objects as nanoscale 
acoustic transducers for hypersonic wave imaging\cite{Amziane2011}. 
Non destructive imaging with nanometric resolution in both depth and 
lateral direction is now one step ahead.
 
Furthermore, there is a recent and growing 
interest in both the electromagnetic and acoustic wave communities for
materials and waveguides that exhibit backward propagating waves\cite{veselago2006left}. 
In backward waves, the phase velocity describing the 
propagation of individual wave fronts in a wave packet and 
the energy flux of the wave, characterized by the Poynting vector are 
anti-parallel. This anti-parallel propagation constitutes the 
definition of a negative index material and opens the way to 
a large variety of intriguing physical 
phenomena\cite{veselago1968electrodynamics}. There now exist  
multiple experimental evidences of this effect, and 
applications of negative index materials for electromagnetic waves  
have already been conceived\cite{Parazzoli2004}. The possibility of such 
backward wave motion in elastic waveguides has been revealed by the early work of 
Lamb\cite{lamb1904group}. Many different elastic waveguides such 
as cylinders\cite{meitzler1965backward} or 
surface acoustic waves\cite{Maznev2009} can also exhibit backward wave motion. 
Recently, negative refraction and focusing of elastic Lamb 
waves have been investigated but are still confined to the low-MHz range\cite{Philippe2015,Bramhavar2011}.
In this letter we provide evidence of backward wave propagation in
one-dimensional elastic waveguide in the gigahertz range. We use 
time-resolved pump-probe spectroscopy with spatially shifted pump and 
probe beams to excite and detect the backward wave motion. As a first 
application, unambiguous mode identification using backward wave 
propagation is investigated.

\begin{figure}[!ht] 
\includegraphics{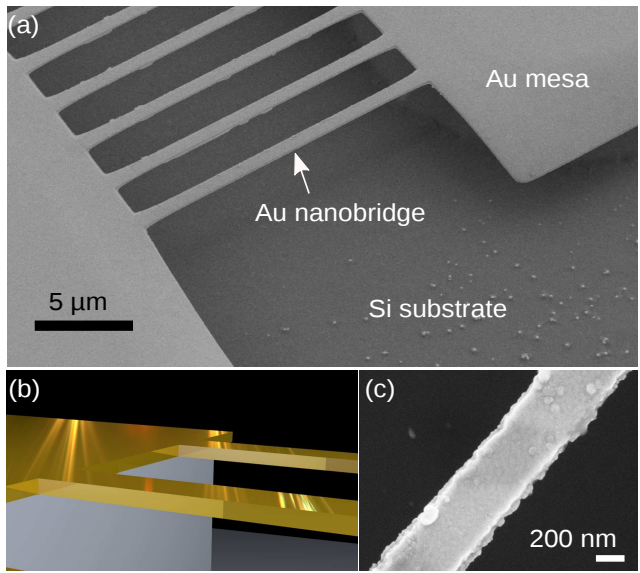}
  \caption{(a) Scanning electron microscope image (top view) of several 
gold nanobeams suspended a few micrometers 
above the silicon substrate. (b) Artist view of nanobeams with a rectangular 
cross section suspended above the silicon substrate. (c) Scanning Electron
 microscope image (top view) of a $\SI{400}{\nano\metre}$ wide gold nanobeam. 
 The edge roughness is lower than $\SI{20}{\nano\metre}$.}
\label{fig1}
\end{figure}  

In order to study guided modes in one dimensional elastic 
waveguides, gold nanostructures with a rectangular cross section are 
conceived (Fig.\ref{fig1}). Their thickness, $h$, is fixed to 
$h=\SI{110}{\nano\metre}$ and their width, $W$, ranges from $200$ to 
$\SI{800}{\nano\metre}$. As shown in Fig.\ref{fig1}(a), 
the gold stripes are connected to large mesas 
at both ends. To reduce the influence of the silicon 
substrate on the elastic confinement, the gold nanostructures are 
suspended several micrometers above the substrate as illustrated in 
Fig.\ref{fig1}(b). This geometry results in what we call ``gold 
nanobeams''. All boundaries are thus mechanically free. 
As their length is greater than 
$\SI{15}{\micro\metre}$, they are considered as 
infinitely long and the fixed boundary conditions at the
extremities can be neglected. 
The samples are fabricated by electronic lithography coupled with 
anisotropic wet 
etching on silicon substrates. To prepare these structures, single face 
polished Si(001) substrates are used. First, an electronic resist 
(PMMA-950K-A6) is spin coated on the Si wafer surface. The exposition 
is made at 20kV using a Zeiss Supra $40$ scanning electron microscope 
equipped with a Raith Elphy Quantum module. After being developed, rinsed 
with water and blown dry, the sample is coated with 
$\SI{10}{\nano\metre}$ of chromium to improve the adhesion, followed 
by a $\SI{110}{\nano\metre}$ thick gold layer. Then, a lift-off process 
in aceton with careful ultrasonic agitation is performed to 
delimit the future gold nanobeams. Finally, the sample is
dipped in a hot $40\%$ KOH solution to partially etch the Si 
substrate and free the desired gold nanobeams. 
This quick anisotropic Si etching process leads to the formation of 
suspended gold nanobeams between two big gold pads. 
The key point of this proces is that the main axis of the beams has been rotated 
$\ang{45}$ with respect to the $<110>$ direction of Si substrates 
to allow this underetching. 

\begin{figure}
  \includegraphics{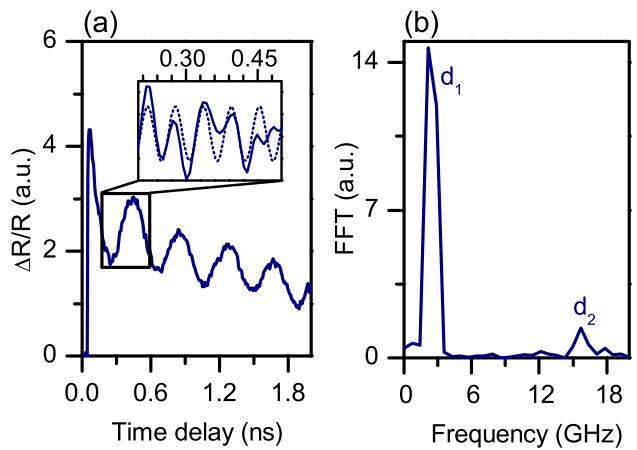}
  \caption{(a) Typical time-resolved modulation of the reflected probe beam 
$\Delta R/R$ of a $\SI{400}{\nano\metre}\pm \SI{20}{\nano\metre}$ wide and 
$\SI{110}{\nano\metre}$ thick gold nanobeam that exhibits a low frequency 
oscillation at $\SI{2.8}{\giga\hertz}$ superimposed to a higher modulation 
around $\SI{16}{\giga\hertz}$ (the inset is a zoom where the lower modulation 
is removed, dotted line is a $\SI{16}{\giga\hertz}$ sinus). (b) Fourier 
Power spectra of the time transient signal (a) showing both $\SI{2.8}{\giga\hertz}$
 and $\SI{16}{\giga\hertz}$ acoustic signatures. $d_1$ and $d_2$ refer to 
the first dilatational modes.}
\label{fig2}
\end{figure} 

\begin{figure}[!h]
	\includegraphics{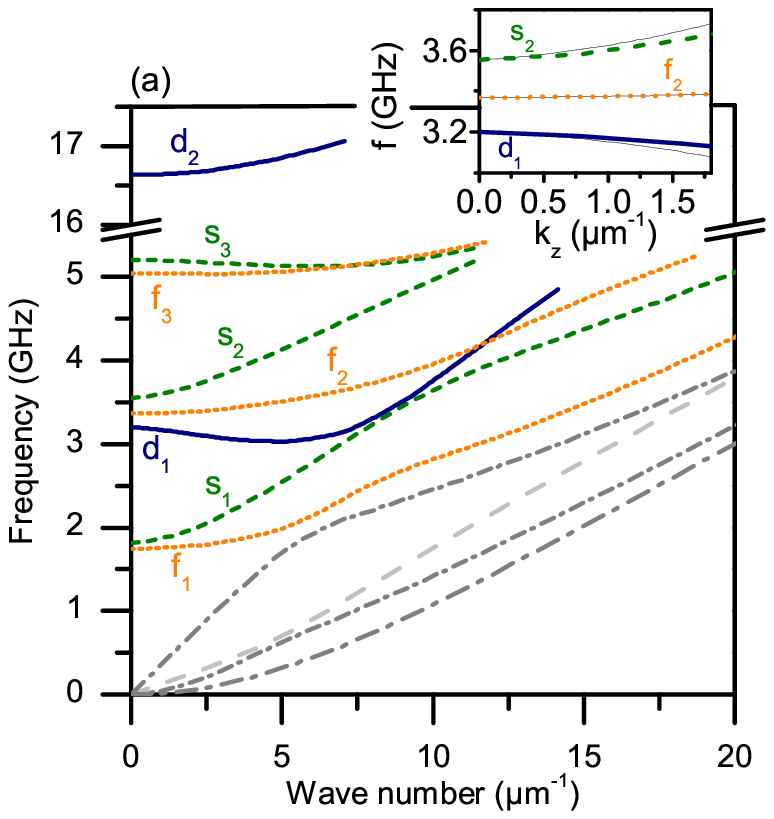}
	\begin{flushleft}
    \includegraphics{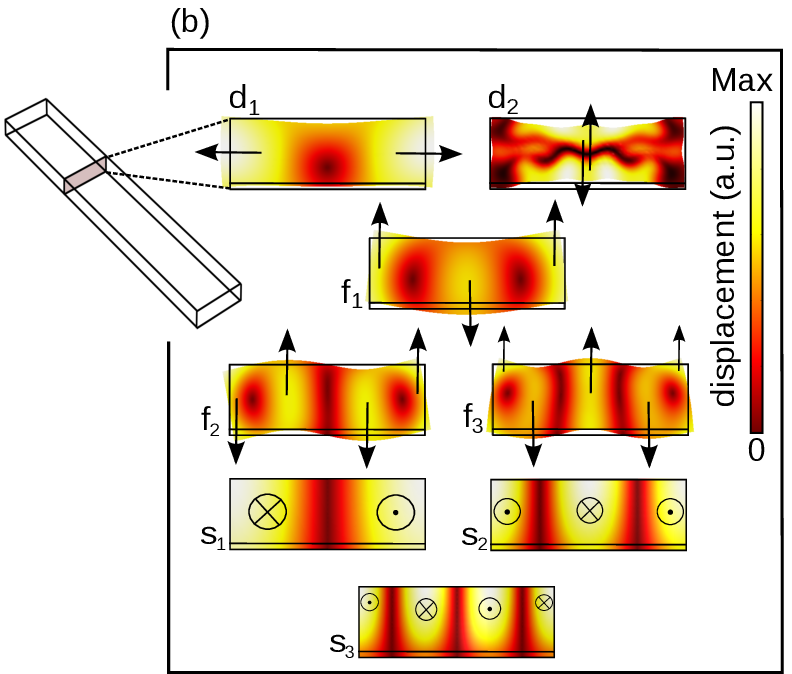}
    \end{flushleft}
  \caption{(a) Acoustic modes dispersion relations in a 
  $\SI{370}{\nano\metre}$ wide and $\SI{110}{\nano\metre}$ thick gold 
  nanobeam (with a $\SI{10}{\nano\metre}$ thick 
adhesion layer of chromium). $f_i$, $s_i$ and $d_i$ refer respectively 
to flexural, shear and dilatational modes. The inset shows the parabolic fit of 
three specific modes at low wave number used to evaluate the 
corresponding propagating wave packet. The parabolic fit is 
  respectively $f_{d_1}(k) = -3.4 \times 10^{-2}\,k^2 - 9.2 \times 10^{-5} \,k + 
  3.2~\mathrm{GHz}$, $f_{s_2}(k) = 5.0 \times 10^{-2}\,k^2 + 5.0 \times 
  10^{-4}\,k + 3.6~\mathrm{GHz}$ and 
  $f_{f_2}(k) = 5.7 \times 10^{-3} \,k^2 +2.8 \times 10^{-4}\,k + 
  3.4~\mathrm{GHz}$ for the modes $d_1$, $s_2$ and $f_2$ with the wave 
  number $k$ in $\SI{}{\micro\metre}^{-1}$. (b) Cross-sectional 
representation of the displacement field of 
three flexural ($f_1$,$f_2$ and $f_3$), two dilatational 
($d_1$ and $d_2$) and three shear ($s_1$,$s_2$ and $s_3$) acoustic 
modes. The black arrows show the displacement direction.}
	\label{fig3}
\end{figure}

Our experimental setup working in reflection geometry was described 
in detail elsewhere\cite{Bienville2006}. Ultrafast pump-probe spectroscopy 
experiments are performed using a mode-locked Ti:sapphire 
(MAI-TAI Spectra) laser source operating at $\SI{800}{\nano\metre}$ with
a pulse duration below $\SI{100}{\femto\second}$ at the laser output 
and a repetition rate of 
$\SI{78.8}{\mega\hertz}$. Synchronous detection on the sample 
reflectivity is performed by 
modulating the pump beam at $\SI{1.8}{\mega\hertz}$. Both pump and 
probe beams are focused by means of a microscope objective 
with a $NA = 0.9$ and are 
normally incident on the sample. The laser spots can be focused around 
$\SI{1}{\micro\metre}$ diameter at $1/e^2$. 
A telescope is fixed on a XY piezoelectric 
stage such that the probe beam is laterally positioned with respect to the 
fixed pump beam. A two-color 
experiment is performed by doubling the pump frequency 
($\lambda = \SI{400}{\nano\metre}$) with a nonlinear cystal (BBO)
to avoid scattered light coming from the pump beam. 
A dichroic filter located
in front of the diode system suppresses the light of the pump beam, 
its power is reduced around $\SI{500}{\micro\watt}$ and the power of the 
probe beam does not exceed $\SI{15}{\micro\watt}$. 

\begin{figure}
  \includegraphics{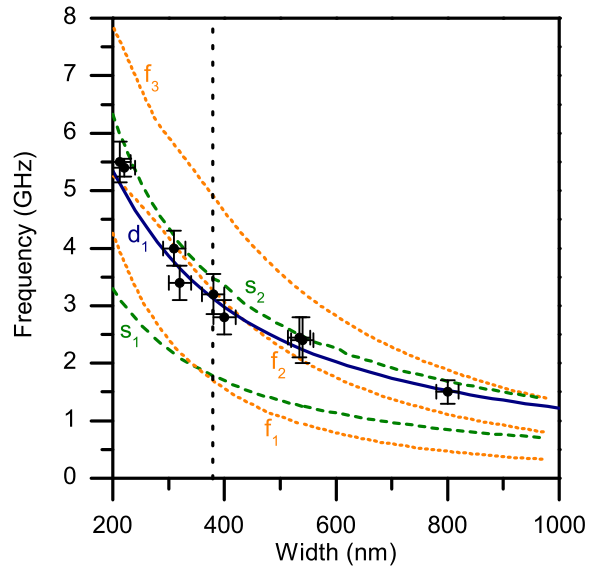}
  \caption{Using a finite elements method, the acoustic mode frequency at 
  wave number $k=0$ is plotted versus the nanobeam width. Flexural, 
  shear and dilatational modes are respectively the orange dotted lines, 
  the green dashed lines and the blue solid lines. Experimental 
  measurements are the full black circle with bars. The vertical 
  black dashed line at $\SI{370}{\nano\metre}$ is the specific width where 
  dispersion relations have been calculated (Fig.\ref{fig3}) and 
  propagation has been observed (Fig.\ref{fig5}).
   }
  \label{fig4}
\end{figure} 

\begin{figure}[!ht] 
\includegraphics{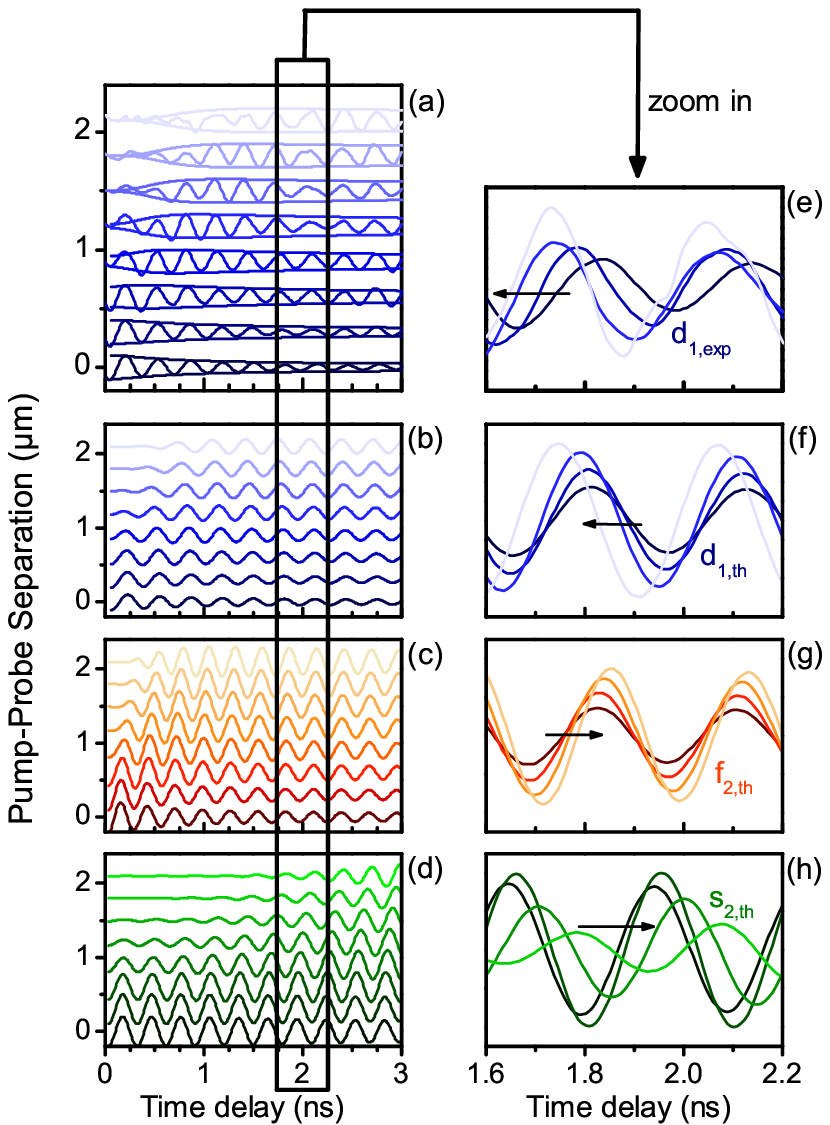}
  \caption{(a) Experimental transient relative reflectivity signal 
  at different spatial pump-probe separations (the decreasing exponential 
  background has been removed). The offset 
  corresponds to the spatial pump-probe shift and the reflectivity
  modulation is normalised by the maximum. The envelope of the eight experimental traces 
  in (a) are the simulated envelope of the first dilatational mode 
  $d_1$ after propagation . (b), (c), (d) are the simulated transient relative reflectivity signal 
  of the modes $d_1$, $f_2$ and $s_2$ according to the parabolic fit
  of the dispersion relation (Fig.\ref{fig3}(a)). 
  (e), (f), (g), (h) are respectively the same signals as in (a), (b), (c), 
  (d) zoomed in a shorter time window. Only half of the eight curves are 
  plotted for readability. The offset is also suppressed in order to 
  observe the propagation direction of the phase of the wave. The black 
  arrow indicates this propagation direction.}
  \label{fig5}
\end{figure} 

Such experimental 
conditions place us in the thermoelastic regime and the acoustic signal and
 optical reflectivity remain stable during all the average processing.
The reflectivity from the sample is measured by an avalanche photodiode and 
analyzed with a lock-in amplifier. A maximum pump-probe
time delay equal to $\SI{12}{\nano\second}$ is achieved using a mobile reflector 
system mounted on a translation stage. 

First, let us consider the case where the pump and probe beams are 
spatially superimposed. Fig.\ref{fig2} shows the time domain reflectivity 
change measurements $\Delta R/R$ obtained on a $\SI{400}{\nano\metre}\pm \SI{20}{\nano\metre}$ wide 
(determined by SEM measurement) and 
$\SI{110}{\nano\metre}$ thick gold nanobeam in a usual reflectometry set-up. 
After a sharp and sudden rise, 
a slowly decreasing non-oscillatory background in $\Delta R/R$ is observed. 
This is the signature of the rapid heating of the electron gas 
induced by the pump pulse absorption, followed 
by the slow cooling down process. This thermal 
stress launches the acoustic vibrations of the 
nanobeam\cite{thomsen1984coherent}. The time-resolved signature 
also shows a superposition of high
(inset Fig.\ref{fig2}(a)) and low frequency oscillations. By 
performing a numerical fast Fourier transform (Fig.\ref{fig2}(b)) 
or by fitting the time oscillation with a damped sinus function, one can extract 
a $\SI{2.8}{\giga\hertz} \pm \SI{0.5}{\giga\hertz}$ and a $\SI{16}{\giga\hertz}
\pm \SI{0.5}{\giga\hertz}$ acoustic signatures. In the 
following we consider that the Young modulus, the Poisson's ratio and the 
density of gold are respectively $\SI{79}{\giga\pascal}$, $0.44$ and 
$\SI{19300}{\kilo\gram.\per\cubic\metre}$\cite{meyers2009mechanical}. The longitudinal 
$v_L$ and transverse $v_T$ sound velocities are then 
$\SI{3.6}{\micro\metre\per\nano\second}$ and $\SI{1.2}{\micro\metre\per\nano\second}$ respectively. Thus, the 
higher $\SI{16}{\giga\hertz}$ frequency appears to be the thickness vibration 
of the beam owing to the fact that $v_L/(2 h) = \SI{16.5}{\giga\hertz}$. 
The displacement field of this $d_2$ eigenmode is plotted in Fig.\ref{fig3}(b). 
At this point, we verify that this thickness resonance signature is 
also observed when the pump-probe experiment is undertaken on the 
gold mesa using an interferometric scheme in this case 
as photoelastic signals are negligible in gold layers\cite{Perrin1999}. 
Consequently, the 
reflectometry measurement undertaken on the gold nanobeam is probably 
partly an interferometric like detection with the light reflected by the 
silicon substrate underneath acting as a reference mirror. 

The case of the lower 
frequency $\SI{\sim3}{\giga\hertz}$ is dealt with by performing an eigenmode 
analysis by finite elements method. The dispersion relation of a 
$\SI{370}{\nano\metre}$ wide nanobeam is plotted in Fig.\ref{fig3}. Several 
acoustic modes with non-zero cut-off frequencies are revealed. Three 
distinct families emerge : the flexural modes $f_1$, $f_2$ and 
$f_3$, the dilatational modes $d_1$ and $d_2$ and the shear 
modes $s_1$, $s_2$ and $s_3$ (see Fig.\ref{fig3}(b)). 
The frequency evolution of $s_1$, $s_2$, $f_1$, $f_2$, $f_3$ and $d_1$ 
with the nanobeam width is plotted in Fig.\ref{fig4}. 
The measured frequencies on different gold nanobeams are also 
reported. Owing to the measurement uncertainties, 
the experimental frequencies ranging from $\SI{1.7}{\giga\hertz}$ to 
$\SI{5.5}{\giga\hertz}$ may be consistent with the
frequencies of the modes $f_2$, $s_2$ and $d_1$ which exhibit close 
cut-off frequencies. Consequently, additional evidence on the detected mode is 
needed. Given that the excitation process imposes the relative amplitudes 
of the excited modes, one can use the initial displacement field 
projection on the orthogonal basis formed by the nano-object eigenmodes 
to identify which mode will be efficiently excited\cite{Hu2003,Crut2011}. 
Here we propose to discriminate unambiguously between the different acoustic 
modes by probing their propagation. Very different behaviors 
are expected owing to the very different slopes of the dispersion 
relationship.

In order to be sensitive to the propagation phenomenon, the pump and probe beams
have to be separated\cite{Kelf2011}. The transient reflectivity 
measured on a $\SI{370}{\nano\metre}$ width gold nanobeam, with 
pump-probe separation ranging from $\SI{0}{\micro\metre}$ to 
$\SI{2.1}{\micro\metre}$, is presented in Fig.\ref{fig5}(a). The vertical scale
corresponds to the pump-probe spatial separation. The exponential 
thermal background has been removed. As physical insight on the 
attenuation process is beyond the scope of this study, each 
transient reflectivity signal is normalized by its maximum : the eighth 
experimental signal with  
$\SI{2.1}{\micro\metre}$ spatial separation is one order of magnitude less 
intense than the first one when pump and probe beams are spatially superimposed.
The signal is mainly composed by a $\SI{3.2}{\giga\hertz}$ oscillation. As the 
pump-probe separation increases, this wave-packet shifts to a
longer time delays as expected for a propagation guided along the 
nanobeam. This is analogous to what has already been observed in 
copper nanowires\cite{jean2014direct}. However, let us have a closer look 
at the individual wave fronts (Fig.\ref{fig5}(e)). It appears that 
as the wave-packet propagates (from darker to lighter curve), the individual 
wave fronts move to smaller time-delay. To investigate this propagation 
quantitatively, we use the previously developed 
expression of the reflectivity for a propagating gaussian wave-packet 
with the parabolic dispersion relation 
$\omega(k)=\alpha k^2 + \beta k + \gamma$\cite{jean2014direct}

\begin{eqnarray}
\Delta r(t) & \propto & \operatorname{Re}\left(\dfrac{\exp(-\mathrm{i}\gamma t)}
{\sqrt{\sigma^2+32\mathrm{i}\alpha t}} \exp
{\left(-\dfrac{8(z_0-\beta t)^2}{\sigma^2+32\mathrm{i} \alpha t}\right)}\right)
\label{reflec}
\end{eqnarray}

where $z_0$ is the spatial pump-probe separation and $\sigma$
is the root of the sum of squares of the pump and probe beam diameters 
at $1/e^2$. A polynomial fit at small wavenumber for $d_1$, 
$f_2$ and $s_2$ (see inset Fig.\ref{fig3}(a)) is used. We then evaluate the 
theoretical reflectivity signal for these three modes $d_1$ (Fig.\ref{fig5}(b)), 
$f_2$ (Fig.\ref{fig5}(c)) and $s_2$ (Fig.\ref{fig5}(d)) at different pump 
and probe separations. 
The progressive shift of the wave-packet 
is very slow for the $f_2$ mode (Fig.\ref{fig5}(c)), 
due to the very flat dispersion relation. 
Both the $f_2$ and $s_2$ modes show parallel propagation of the individual 
wave fronts and the wave packet (Fig.\ref{fig5}(g),(h)), 
whereas the $d_1$ mode exhibits the expected backward propagation behavior 
(Fig.\ref{fig5}(f)). According to our simplified analytical 
expression, the parallel (forward) or anti-parallel (backward) behavior 
is strongly dependent on the sign of the curvature parameter $\alpha$. The 
first dilatational mode $d_1$ is the only one to exhibit a negative 
curvature $\alpha$ at low wave numbers $k$ which are the only wave 
numbers excited owing to the laser spot diameters of around 
$\SI{1}{\micro\metre}$.
To confront the experiment, the envelope of the theoretical 
signal $d_1$ is also plotted on the experimental oscillation 
and shows a good agreement (Fig.\ref{fig5}(a)). 
Finally, it could be argued that the experimental 
signal is a superposition of two or more acoustic modes. However, 
if a contribution of beyond $10\%$ of $f_2$ or $s_2$ is added 
in the simulated signal $d_1$, the resulting simulated 
propagative wave no longer exhibits the backward wave behavior. The backward 
wave observation in our 1D elastic waveguide is thus the unique 
fingerprint of the first width dilatational mode $d_1$.

In conclusion, two dilatational acoustic modes are investigated on 
e-beam lithographed gold nanobeams by pump-probe time-resolved 
spectroscopy leading to the observation of backward wave 
propagation in the gigahertz range. 
As a first application, we show unambiguous acoustic 
mode discrimination using this unique property in the dispersion 
relations. Furthermore, as already observed in 2D Lamb waveguides, 
one can imagine to tune the width and thickness of our nanobeam, thus opening the way 
to negative refraction physics in acoustics at the nanoscale.

\begin{acknowledgments}

The authors would like to thank Michael Rosticher at the ENS for the EBL
steps during the maintenance in our facilities, M\'elanie Escudier for the 
gold deposition process and Frances Edwards for her careful proofreading of the 
manuscript's English. 

\end{acknowledgments}

%----------------------------------------------------------------------------------------
%	BIBLIOGRAPHY
%----------------------------------------------------------------------------------------

\label{Bibliography}

\bibliographystyle{apsrev4-1} % Use the "unsrtnat" BibTeX style for formatting the Bibliography

\bibliography{bibliographie} % The references (bibliography) information are stored in the file named "Bibliography.bib"

%----------------------------------------------------------------------------------------

\end{document}